# Multi-Marker Similarity enables reduced-reference and interpretable image quality assessment in optical microscopy.

Elena Corbetta[1,2], Thomas Bocklitz[1,2]

1. *Leibniz Institute of Photonic Technology, Member of Leibniz Health Technologies, Member of the Leibniz Centre for Photonics in Infection Research (LPI), Albert-Einstein-Strasse 9, 07745 Jena, Germany.*
2. *Institute of Physical Chemistry (IPC) and Abbe Center of Photonics (ACP), Friedrich Schiller University Jena, Member of the Leibniz Centre for Photonics in Infection Research (LPI), Helmholtzweg 4, 07743 Jena, Germany*

**Abstract**

Optical microscopy contributes to the ever-increasing progress in biological and biomedical studies, as it allows the implementation of minimally invasive experimental pipelines to translate the data of measured samples into valuable knowledge. Within these pipelines, reliable quality assessment must be ensured to validate the generated results. Image quality assessment is often applied with a full-reference approach to estimate the similarity between the ground truth and the output images. However, current methods often show poor agreement with visual perception and lead to the generation of a variety of full-reference metrics tailored to specific applications. Additionally, they rely on pixel-wise comparisons, which emphasize local intensity similarity while often overlooking comprehensive and interpretable image quality assessment.

To address these issues, we have developed a multi-marker similarity method that compares standard quality markers, such as resolution, signal-to-noise ratio, contrast, and high frequency components. The method computes a similarity score between the image and the ground truth for each marker and then combines these scores into an overall similarity estimate. This provides a full-reference estimate of image quality while extracting global quality features and detecting experimental artifacts. Multi-marker similarity provides a reliable and interpretable method for image quality assessment and the generation of quality rankings. By focusing on the comparison of quality markers rather than direct image distances, the method enables reduced reference implementations, where a single field of view is used as a benchmark for multiple measurements. This opens the way for a reliable and automatic evaluation of big datasets, typical of large biomedical studies, when manual assessment of single images and the definition of the ground-truth for each field of view is not feasible.

# Introduction

The advancement of biological and medical studies is made possible by the progress in multi-disciplinary fields, including the development of experimental techniques that allow a reliable collection of large datasets while minimizing sample degradation and enabling an accurate processing and analysis of the collected data. In this direction, robust experimental workflows are essential to ensure repeatable results and a reliable diagnosis. Among the established experimental techniques for biomedical applications, optical microscopy includes a family of methods that allow minimally invasive measurements. Optical microscopy emerged as a robust diagnostic tool thanks to the possibility to implement reliable experimental pipelines to translate the experimental measurement into valuable knowledge. To achieve this goal in image-based studies, researchers must implement reliable pipelines to select a good image from the measurement session, enhance the quality of the image, and extract quantitative knowledge (Figure 1 (a)). The first step of the pipeline is the acquisition of raw data by utilizing experimental techniques to measure the sample. Then, reconstruction and processing steps are implemented to generate an image from the raw data, if necessary, and improve the image quality by removing experimental artifacts. Finally, the image can be translated into quantitative information by image analysis algorithms. These steps are interconnected, and the success of each process depends on the quality of the output of the previous step. For this reason, image quality assessment (IQA) must be integrated in the experimental pipeline to tune the algorithms and validate each intermediate result before moving to the next task. Current microscopy-based studies can generate hundreds of images, which cannot be inspected manually. Therefore, automatic IQA methods are required within every research study for the selection of the good data and FAIR data management.[1-5]

Depending on the data availability and the knowledge of the experimental process, IQA can be performed choosing from a wide variety of approaches. Irrespective of the implementation, IQA should be objective, reliable, automatized and interpretable. IQA methods are commonly classified into three categories, depending on the knowledge necessary for their implementation: no-reference (NR), reduced-reference (RR), and full-reference (FR) IQA. These three approaches require, respectively, no prior high-quality data for comparison, limited knowledge about the characteristics of the ideal image, or the definition of the ground-truth (GT), which is used for comparison (Figure 1 (b)).[6-8] This manuscript targets FR IQA methods and provides specific strategies to reduce the amount of required information required and move to a RR implementation. Despite the high amount of prior knowledge required, FR IQA is a powerful method to validate experimental techniques, to benchmark new algorithms, and to implement supervised processes. In this context, FR IQA can be implemented with known datasets, for which the GT can be easily defined. However, FR IQA is less feasible in real case scenarios where the output of a measurement is not known a priori. For this reason, it is important to explore RR IQA, where possible.[9,10]

FR IQA is based on the definition of some sort of similarity or distance between the GT and the image to be analysed. In the current state of the art, many FR metrics have been adapted from natural image studies and applied to optical microscopy studies. A widely used similarity metric is the structural similarity index (SSIM)[6], which ranges from -1 and 1 and is maximized for perfect similarity between two images. SSIM is composed of three factors, which compute local similarity between two images in terms of luminance, contrast, and structure. Other FR metrics define a mathematical distance between two images: they include the peak signal-to-noise ratio (PSNR), the mean squared error (MSE), and the mean absolute error (MAE). A third category of FR metrics is represented by correlation measures, such as the Pearson's correlation coefficient (PCC).[11,12]

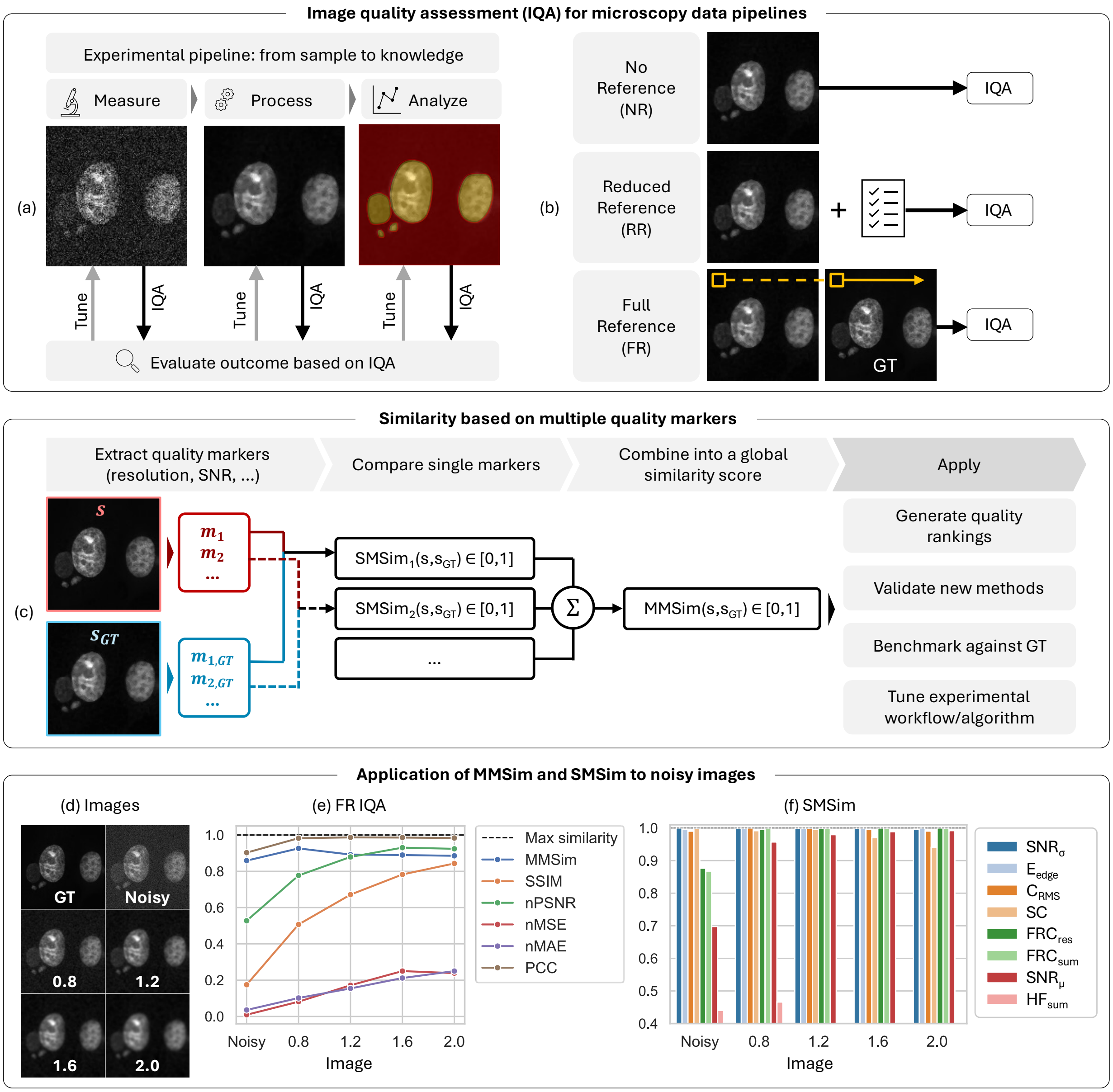

*Figure 1 – **Image quality assessment (IQA) for fluorescence microscopy: introduction to full-reference (FR) metrics and multi-marker similarity (MMSim).** (a) Experimental pipeline for fluorescence microscopy: the measurement, processing and analysis steps are implemented to enhance the image quality and retrieve quantitative information from the measured sample. IQA must be applied at each stage to evaluate the outputs and tune the experimental acquisition, the choice of the processing algorithms, as well as the related hyperparameters. (b) Different approaches for IQA: from the top, no-reference (NR) metrics provide a direct evaluation without the use of reference data, reduced-reference metrics compare the image with a set of known requirements, full-reference metrics compare the image with the Ground Truth (GT), computing a pixel-wise or a sliding-window-wise comparison. (c) Schematics for the computation of multi-marker Similarity (MMSim). The method is not based on pixel-wise intensity comparison, but on the extraction of global quality markers from the image data (s) and the GT ($s_{GT}$). Single markers are compared individually by computing single-marker similarity (SMSim) scores, and then they are averaged into a global MMSim score. (d) Simple example application of MMSim to a denoising problem. GT is the reference image, Noisy is the image with mixed Poisson-Gaussian noise, and the following images are denoised by Gaussian filtering with standard deviation (in pixel) as reported in the labels. (e) MMSim compared with five state-of-the-art metrics: Structural Similarity Index (SSIM), Peak Signal-to-Noise Ratio (PSNR), Mean Squared Error (MSE), Mean Absolute Error (MAE), Pearson Correlation Coefficient (PCC). nPSNR, nMSE and nMAE are normalized to be bound to 0-1 for a comparable visualization with the other FR metrics. (e) SMSim for 8 different quality markers applied to the images. SMSim are plotted from the less-varying score (blue, $SNR_\sigma$) to the one with highest variation in the dataset (pink, $HF_{sum}$).*

All the mentioned FR metrics rely on intensity comparisons computed between single pixels or small sliding windows. In many cases, this local approach results in poor agreement with visual perception and difficult interpretation of the final score. This was underlined by previous studies on perceptual image resolution, which also highlighted that individual FR metrics show a different behaviour for different image degradations and are more sensitive to specific image artifacts.[13,14] Many alternative implementations of state-of-the-art metrics have been proposed to address these drawbacks in specific case studies. For example, different replacements have been implemented for SSIM. Multi-scale SSIM (MS-SSIM) has been proposed to take into account the different scale of imaged structures[15]; other variants have been implemented to address the poor agreement of SSIM with visual perception in presence of blurred and noisy images[16], and to adapt it to the different dynamic range, noise content and offset of microscopy images compared to natural images[17]. For example, a different offset of a high-quality image compared to the GT will result in a strong change in luminance and an increased distance in intensity, leading to a lower quality score regardless of the actual quality and similarity content. All the issues mentioned above result in a lack of interpretability and reliability of state-of-the-art metrics in some applications and leaves a desire for a more comprehensive IQA.[3,8,13,14,18]

To tackle this challenge, we introduce the multi-marker similarity (MMSim), a new approach for FR IQA that generates a similarity score from multiple quality markers extracted from the images and from the GT. As schematized in Figure 1 (c), the workflow for MMSim starts with the computation of multiple NR quality metrics for the image under analysis and the GT, such as resolution, signal-to-noise ratio, contrast, structural complexity, and so on.[19] Then, a single-marker similarity (SMSim, in the figure) score is generated for each marker. Finally, the single-marker similarities are merged into a final overall quality score of the image.

MMSim offers two advantages that are particularly important for FAIR data management of large datasets. Firstly, thanks to this implementation, MMSim is interpretable because it can be fragmented into the individual single-marker similarity scores, which give further insights on the origin of the similarity. In addition, we demonstrate that MMSim can be applied with a RR implementation because, being based on the comparison of quality markers rather than pixel-wise similarities, it allows target features to be extracted from a single GT and used as references for multiple measurements. Therefore, MMSim is particularly advantageous when a collection of images cannot be assessed manually, nor is a GT available for each field of view. In this scenario, a single GT can be defined a priori for each measured structure, and the evaluation result can be further interpreted by tracking the behaviour of the single markers. The MMSim allows the quality comparison of images with different image content, which has been demonstrated in four examples in this manuscript, ranging from the hyperparameter optimization of denoising algorithms to the determination of the optimal focal position of experimental measurements.

# Results

## Single-marker similarity and multi-marker similarity (MMSim) working principle

As introduced above, MMSim falls in the category of FR IQA methods and is based on the extraction of multiple quality markers from the images. Here, we demonstrate briefly the working principle of MMSim and the differences with state-of-the-art approaches. Let's assume to evaluate an image $s$ against the ground $s_{GT}$ by following the workflow of Figure 1 (c). The first step is the computation of the quality markers $m_i$, which are a limited set of NR metrics that have been proved to be reliable markers for the differentiation of a variety of experimental artifacts.[19] Indeed, experimental artifacts generate multiple image degradations that must be detected using a set of parameters. In particular, the default computation of our method includes eight quality metrics: the Fourier ring correlation (FRC), that provides a resolution estimation and the correlation sum, the signal-to-noise ratio, the signal-to-average ratio, the structural complexity, the sum of high frequency components of the image power spectrum, the edge energy ratio and the root-mean-square contrast. More details about the implementation of the metrics are described in the *Methods* section. As we demonstrate later in this manuscript, this set of metrics can be reduced to optimize the computational time by utilizing only the markers that are relevant to the current image evaluation problem. Once the single markers are extracted by $s$ and $s_{GT}$, every pair is utilized to generate a single-marker similarity score, that is symmetric, bound to 0 and 1, and equal to 1 only if $m_i$ and $m_{i,GT}$ are equal. In the last step, the single-marker similarities are averaged into a global quality score, namely the MMSim. For each result of this manuscript, MMSim is compared with the state-of-the-art metrics mentioned above (SSIM, PSNR, MSE, MAE, PCC).[6] More details about the workflow are reported in the *Methods* section.

The bottom panel of Figure 1 shows an example application of our method to a simple denoising problem. Figure 1 (d) shows a noisy image, the related GT, and a set of denoised versions obtained by applying Gaussian filtering with increasing standard deviation to the noisy image. We applied MMSim and other FR metrics to evaluate which is the best denoised option, as shown in Figure 1 (e). The metrics are normalized to have value 1 for maximum similarity. The first relevant observation is that the trend of the metrics is not identical, a problem that has been reported frequently in the literature and that underlines the lack of reliable IQA methods. Most of the state-of-the-art metrics assign a higher score to stronger Gaussian filtering, which generates smoothed images with loss of the detailed sample structure. For all the state-of-the-art metrics, the most smoothed image scores a significantly higher quality than the noisy image. The most smoothed image shows a darker background and noise removal, but also loss of high frequency features that are beneficial for the interpretation of the content. On the other hand, MMSim assigns a comparable score to the noisy image and the strongly smoothed ones and selects the first denoised version as the best one. From a visual point-of-view, this choice seems the most reliable for further interpretation of the image content. One advantage of MMSim is the possibility of further inspection by plotting the single-marker similarity scores. In Figure 1 (f) the single-marker similarity scores are computed for the eight markers and sorted from the less to the most varying in the dataset: the noisy image differs from the GT especially for high frequency components, signal-to-average ratio, and FRC; the first Gaussian filter improves these similarities, while an increasing smoothing lowers again the high frequency similarity and modifies the structural complexity.

# IQA and quality rankings generated by MMSim show good agreement with visual perception

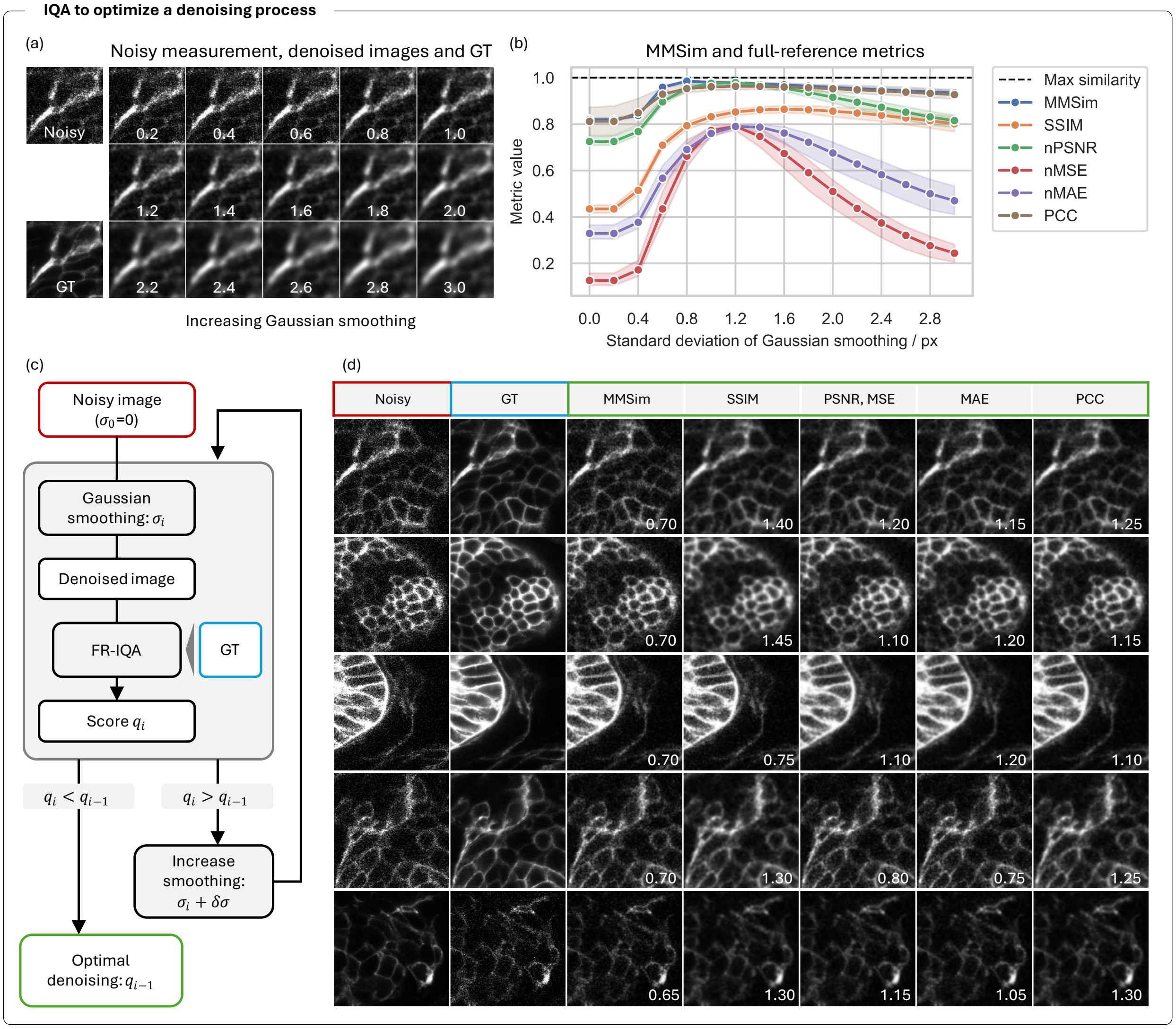

*Figure 2 – **Hyperparameter optimization for denoising with Gaussian filtering.** Five noisy measurements of zebrafish embryos are denoised by Gaussian filters with increasing standard deviation. (a) Example cropped images obtained by filtering one of the five measurements with Gaussian filter. The image labels indicate the standard deviation of the filter. (b) FR metrics averaged over the 5 measurements. nPSNR, nMSE and nMAE are normalized to be bound to 0-1 for a comparable visualization with the other FR metrics. The shaded area is the 95% confidence interval. (c) Schematic of the early stopping algorithm applied to select the best denoised result according to the FR metrics. The process is repeated for every metric, leading to different optimizations. (d) Crops of noisy images, GTs and best images selected by MMSim, SSIM, PSNR, MSE, MAE, PCC. The white labels indicate the standard deviation, in pixel, of the Gaussian filter applied to each denoised image. The result shows that state-of-the-art metrics, especially SSIM and PCC, are biased towards smoothed images, while MMSim favours higher levels of noise, that usually result in a lower loss of detail.*

Here we further investigate the automatic optimization of a denoising output by FR IQA. First, we start by exploring the behaviour of the different metrics for the same filtering problem. Five noisy images of zebrafish embryo are acquired by fluorescence confocal microscopy, and five high-quality images are obtained by average 50 independent noisy measurements with same content.[20] We applied Gaussian filters of increasing standard deviation to the noisy images and used the high-quality images as GT (Figure 2 (a)). Figure 2 (b) shows the result of MMSim and five state-of-the-art metrics averaged over the five measurements for each applied filter. All metrics assign better similarity to smoothed images against the noisy ones and identify the best denoised version within the filtering interval that we selected (i.e., maximum of the curve). PCC, MSE and MAE show higher variability across the evaluation of different measurements. SSIM is characterized by a maximum shifted towards more smoothed versions, as observed in previous studies[16], while MMSim is characterized by an earlier maximum. This trend can be explained by the fact that state-of-the-art metrics rely on a pixel-wise or window-wise comparison of the images, which is more affected by local intensity variations due to noise and might be improved by smoothing.

Given the different behaviour of the metrics, we can build a simple automatic hyperparameter optimization process to select the best denoised output. Figure 2 (c) shows an early stopping algorithm that takes as input a noisy image and applies iteratively an increasing Gaussian filtering. After filtering, FR IQA is computed for the current metric, and the process continues until the similarity is increasing. Figure 2 (d) shows the optimized images selected by the different FR metrics. This result confirms that MMSim selects images with weaker filtering, while the other metrics, especially SSIM, promote stronger smoothing. While sometimes the amount of noise retained by MMSim seems excessive, it is beneficial for the preservation of details of the measurement, that would be otherwise lost with increasing filtering. Indeed, the MMSim score is a compromise between the generation of good resolution, comparable structural complexity, and high frequency details with the GT, and it is not the result of pixel-wise intensity correspondence. Supplementary Figure S1shows the same workflow applied to multi-channel measurements of BPAE cells, with similar outcomes.

In Figure 3 we extend this application to hyperparameter optimization of different standard denoising methods and to the generation of image rankings. As schematized in Figure 3 (a), we apply five denoising methods to the same five measurements processed in the previous example. Each denoising method is applied with five different hyperparameters and FR IQA is computed for each output against the available GTs (one for each measurement, obtained by averaging 50 raw measurements). Figure 3 (b) shows the results obtained for one of the five measurements: the hyperparameters are chosen to provide weakly or almost not denoised images up to images with excessive smoothness and loss of details. Figure 3 (c) shows a subset of FR metrics and MMSim computed for the varying hyperparameters, averaged on the five measurements. The metrics show a comparable trend, but there are differences in the selection of the optimized hyperparameter (labelled in Figure 3 (b) for an easier visualization). On average, MMSim agrees with other metrics and selects images with weaker smoothing, while SSIM is the most biased towards stronger smoothing. SSIM and normalized PSNR show the greatest variability across different samples. On the other hand, MMSim shows a narrow confidence interval. Given that the five measurements are acquired under the same experimental conditions and undergo the same process, this behaviour is a clue that MMSim correctly evaluates the quality of the image independently on the measured sample. In addition, MMSim shows good agreement with visual perception. Further inspection of the MMSim score is reported in supplementary Figure S2, which shows the contribution of the individual single-marker similarities for each denoising method and each hyperparameter.

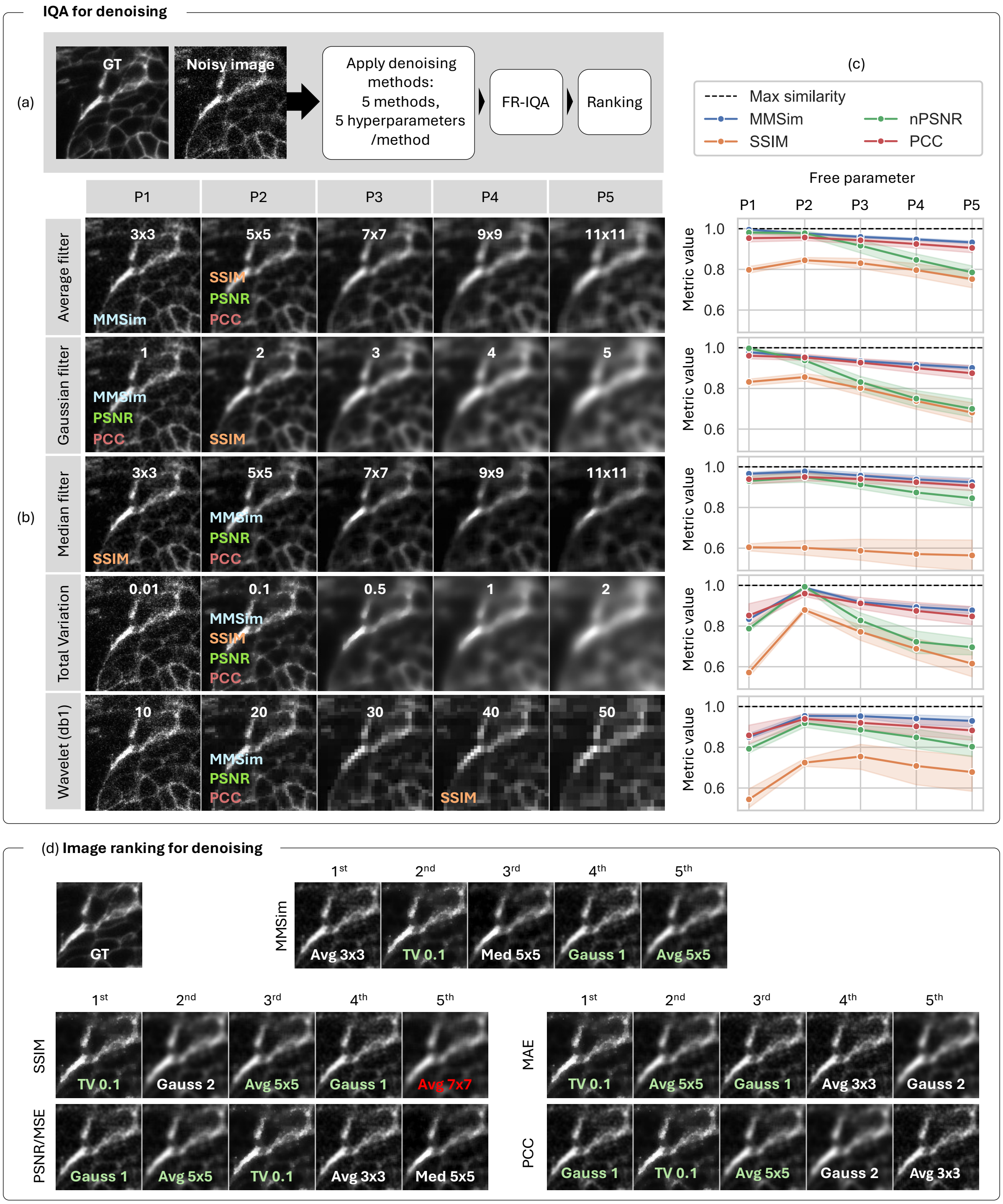

*Figure 3 – **MMSim applied to IQA and quality ranking of denoised images.** (a) Workflow for the generation of the results: a noisy image is corrected by 5 standard denoising methods with a set of 5 parameters for each method. Then, FR IQA is computed, and the quality scores are utilized to rank the images according to their similarity with the GT. (b) Output images processed by five denoising methods (rows) and five hyperparameters per method (columns). Hyperparameters are reported in each cell, and the name of the metrics indicate the best hyperparameter according to each metric: for example, 3x3 px window as best average filter for MMSim, while the other metrics selects the 5x5 px filter. The hyperparameters are window size [px] for average filter and median filter, standard deviation of the 2D Gaussian function for Gaussian filtering, weight for the total variation regularization, and noise standard deviation for wavelet filtering. (c) Trend of four FR metrics, averaged on the 5 samples of the dataset, for the varying hyperparameter of each row. Every sample is evaluated with a specific GT. (d) Top 5 denoised images of the field of view shown in panel (b), selected by the FR metrics. Each cell is labelled with the method and hyperparameter. Red-labelled images appear only once in the top 5 rankings, whereas green-labelled ones appear in all rankings.*

After the evaluation of the denoised images, an overall quality ranking is generated. Figure 3 (d) shows the top five images selected by each FR metrics from the whole dataset of processed images. The labels indicate the applied denoising method and the selected hyperparameter: the labels are white if they occur in at least two rankings, green if they are present in all rankings, and red if they occur only in a single ranking. This result confirms that MMSim is the only metric that avoids the selection of excessively smoothed images among the top 5: indeed, in all the other rankings at least two images appear blurred and with blunted features, which makes the image interpretation more challenging.

## From full-reference to reduced-reference IQA with MMSim

So far, we have demonstrated the successful applicability of MMSim and some scenarios in which our evaluation method is advantageous compared to state-of-the-art metrics. Now we can take a step forward and relax some of the constraints of the IQA process. Since MMSim computes the evaluation by utilizing quality markers, it does not work directly on the input images. Therefore, the use of MMSim can be explored in a RR implementation by comparing different measurements of the same sample against the GT from a single field of view (FOV). The goal of a RR evaluation is to reduce the amount of prior information (i.e., the knowledge of the GT for every FOV), while keeping an IQA result that depends on the quality of the image and not on the specific measured region. Because the other state-of-the-art metrics are based on pixel-wise and window-wise comparison, this RR approach is not usually feasible. In this example we use five measurements of zebrafish embryos indicated as different FOVs in Figure 4 (a). For each FOV, six different measurements with decreasing noise are available, which are obtained by averaging an increasing number of raw noisy images. As before, the GT is obtained by averaging 50 raw images, but here we use only the high-quality image of FOV I (light blue frame) as GT for all FOVs and we compute MMSim and the state-of-the-art metrics. The top plot in Figure 4 (b) shows the metrics against the FOVs, averaged on the six noise levels: MMSim is the only metric independent on the FOV, while all the other scores show strong variability across different FOVs, which is way higher than the variability generated by the noise levels (represented by the confidence interval). The bottom plot shows the evaluation results averaged on the non-reference FOVs (from II to V) and plotted against the noise level: MMSim is the only metric that evaluates the similarity according to the noise level, while most of the other metrics show a flat curve. SSIM is the only state-of-the-art metric showing a clear increasing behaviour with the image quality, but also a strong offset below the maximum score. These conclusions are confirmed by the heatmaps in Figure 4 (c). The cells of the grid are arranged as the images of panel (a), and the colours vary from dark green, assigned to the images with highest similarity with the GT, to dark red, assigned to the images with lowest similarity. On the left, MMSim shows a colour gradient generated mostly by the change in the noise level, with the green images on the right and the red images on the left. SSIM shows a trend that depends on both the noise level and the FOV, while PSNR and PCC are strongly dependent on the image content, with some FOVs featuring the same score across all the noise levels.

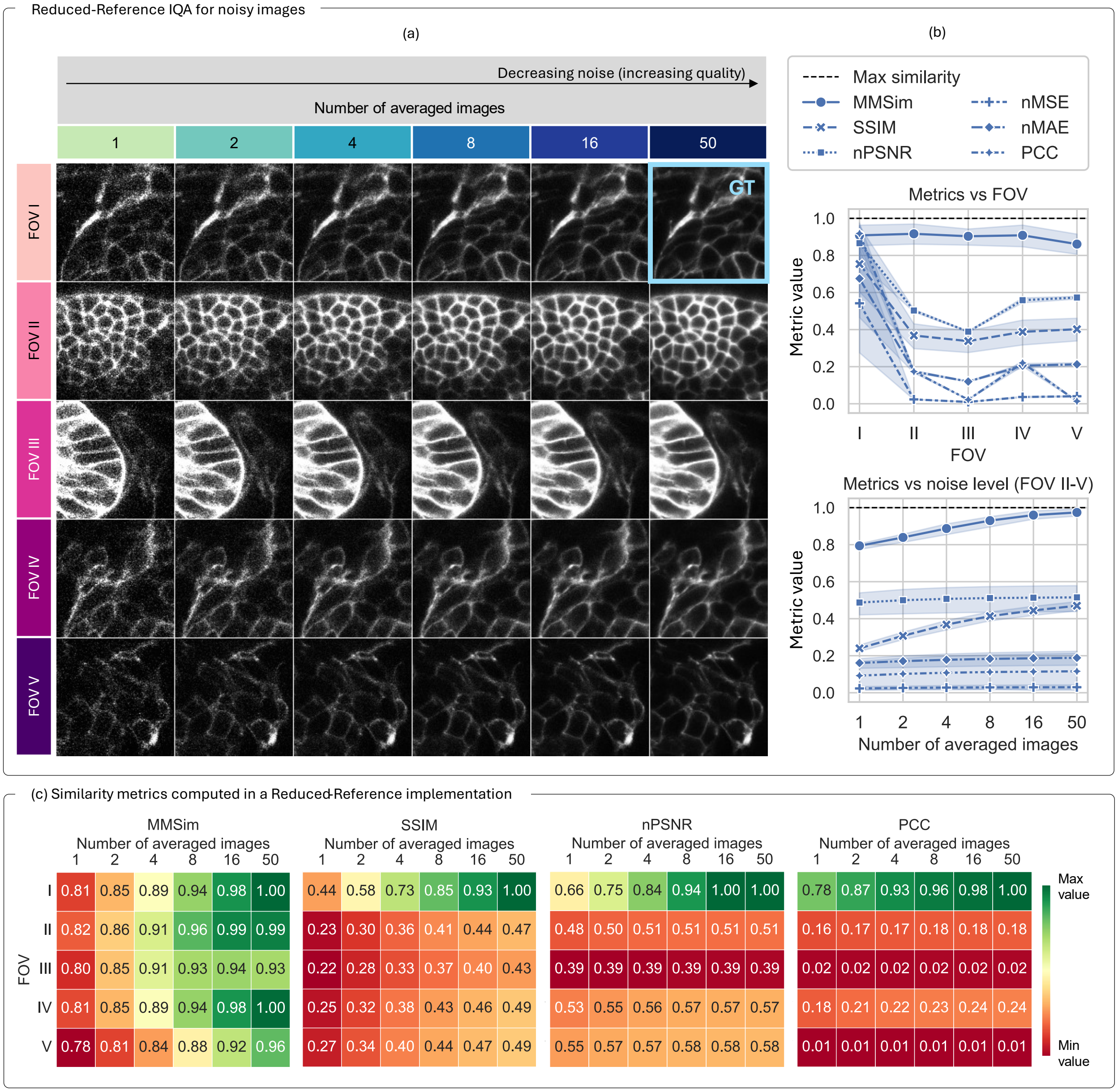

*Figure 4 –* ***From full-reference (FR) to reduced-reference (RR) IQA with MMSim: extraction of quality markers independent on the field-of-view.*** *(a) Experimental measurements of Zebrafish embryos with different levels of noise, obtained by averaging an increasing number of raw images. Each row shows the 6 different levels of noise of a different field of view (FOV). The high-quality measurement of the first FOV (light blue frame) is used as single GT for all the images. (b) MMSim and other FR metrics computed in RR configuration. Top plot: dependence of the metrics on the FOV, averaged over 6 noise levels. Bottom plot: dependence of the similarity metrics on the noise level, averaged on the non-reference FOVs (II-V). The shaded bar is the 95% confidence interval. (c) Heatmaps showing the trend of four similarity metrics computed for the images in panel (a). The colour is adjusted in each heatmap to match the full value range, assigning dark green to the best similarity results and dark red to the largest deviations from the GT. The location of the green cells reveals whether high similarity is assigned to high-quality images of different FOVs (as for MMSim) or to different noise levels of the first FOV (as for state-of-the-art metrics).*

Again, we can generate quality rankings of the full dataset from each metric. Figure 5 shows the rankings plotted as heatmaps, where the colours illustrate two different pieces of information and correspond to the lateral labels in Figure 4 (a): the ranked images are coloured based on their FOV in panel (a) and on the number of averaged images (i.e., the noise level) in panel (b). At the top of both panels, we include a coloured bar, which represents the distribution of the color-coded images for an optimal evaluation that depends only on the image quality: in an ideal scenario, we expect rankings with shuffled FOVs and a high number of averaged images at the top positions. The rankings confirm that MMSim assigns high similarity to images with low noise levels independently on the FOV, while the other metrics show opposite behaviour. Especially in the four bottom rankings, we observe that high quality images are ranked after noisy measurements. This confirms that, among the methods listed here, MMSim is the only approach feasible for a RR evaluation. In supplementary Figure S3 and supplementary Figure S4 we demonstrate that MMSim correlates well with the real noise level of the images and, especially for the RR implementation, it outperforms state-of-the-art methods.

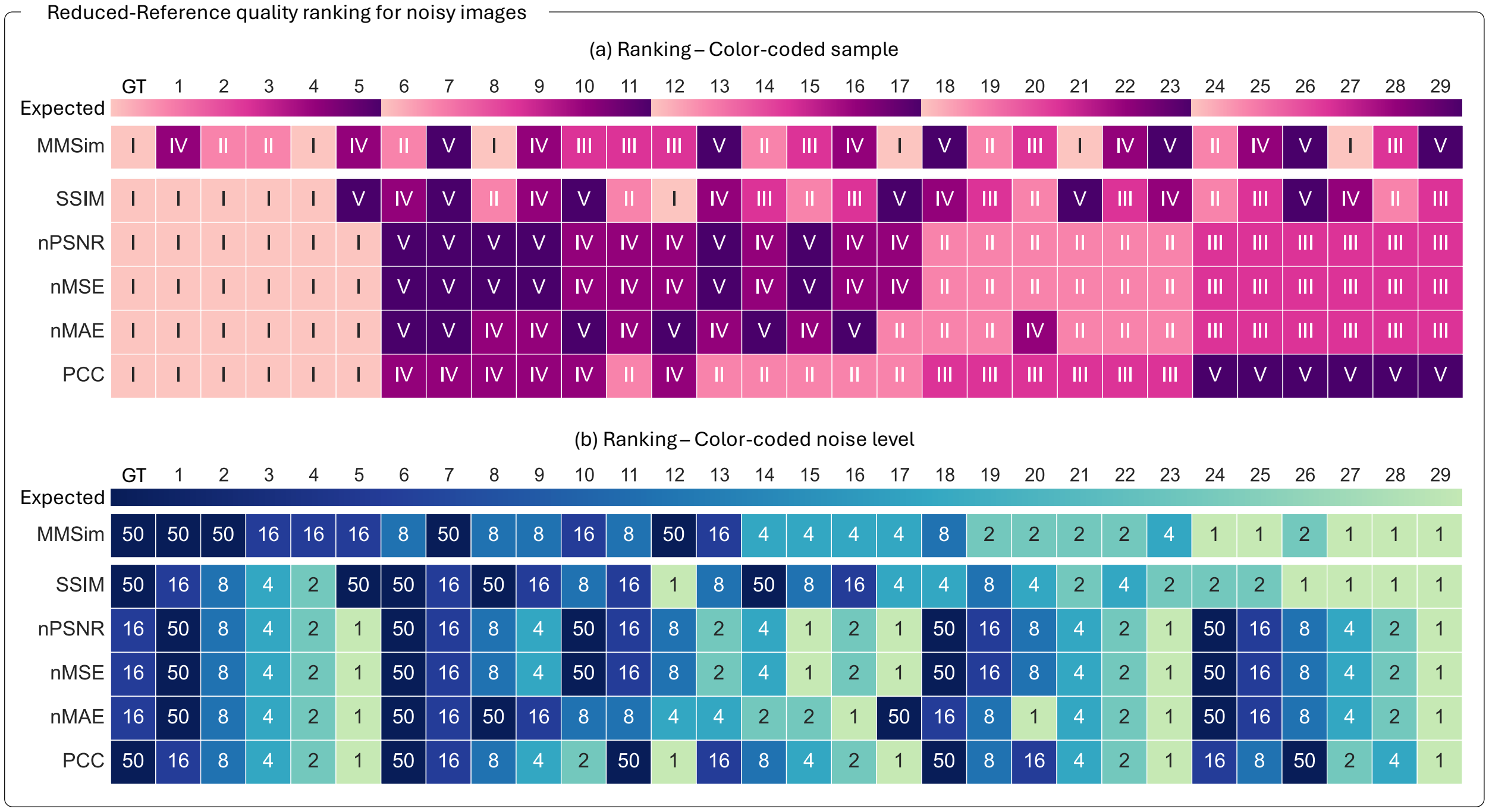

*Figure 5 – **Global quality ranking for measurements of zebrafish embryos (Figure 4) with increasing noise levels.** Each row contains the ranking generated by a different metric in the RR configuration of Figure 4. The same ranking is represented by assigning a different colour to the FOV (panel (a)) and to the noise level (panel (b)). At the top of each panel, a coloured bar (Expected) shows the expected colour distribution of the images in the ranking: for an ideal evaluation, the result should depend only on the image quality, with shuffled FOVs and a high number of averaged images at the top positions. (a) The colour in the ranking indicates the FOV of each image: MMSim ranking is not dependent on the FOV, SSIM ranking is weakly dependent on the FOV, while the other rankings are strongly dependent on the FOV. (b) Same quality ranking as in (a), but the colour indicates the number of averaged images, to show whether the ranking depends on the actual image quality. MMSim has a peculiar behaviour and generates a quality ranking that is strongly FOV-independent and quality-dependent, well aligned with the ideal expected colour distribution.*

# MMSim for reduced-reference selection of the optimal focus

After having explored many applications of MMSim to denoising problems, we apply our method to the automatic selection of the best focal plane in experimental images of BPAE cells nuclei, again with a RR implementation. In addition, we demonstrate an automatic approach to reduce the set of single-marker similarity scores to obtain a reliable evaluation of the images while reducing the computational time. Figure 6 (a) shows image crops extracted from the dataset: eleven FOVs are measured at seven different focal positions (at, above and below the optimal focal plane). The first FOV is taken as reference, and its optimal focal position is utilized as GT for the whole dataset. The figure highlights with yellow frames the optimal focal planes selected by MMSim for the other ten FOVs, with perfect agreement with visual perception, and demonstrates the applicability of our method to a defocusing problem.

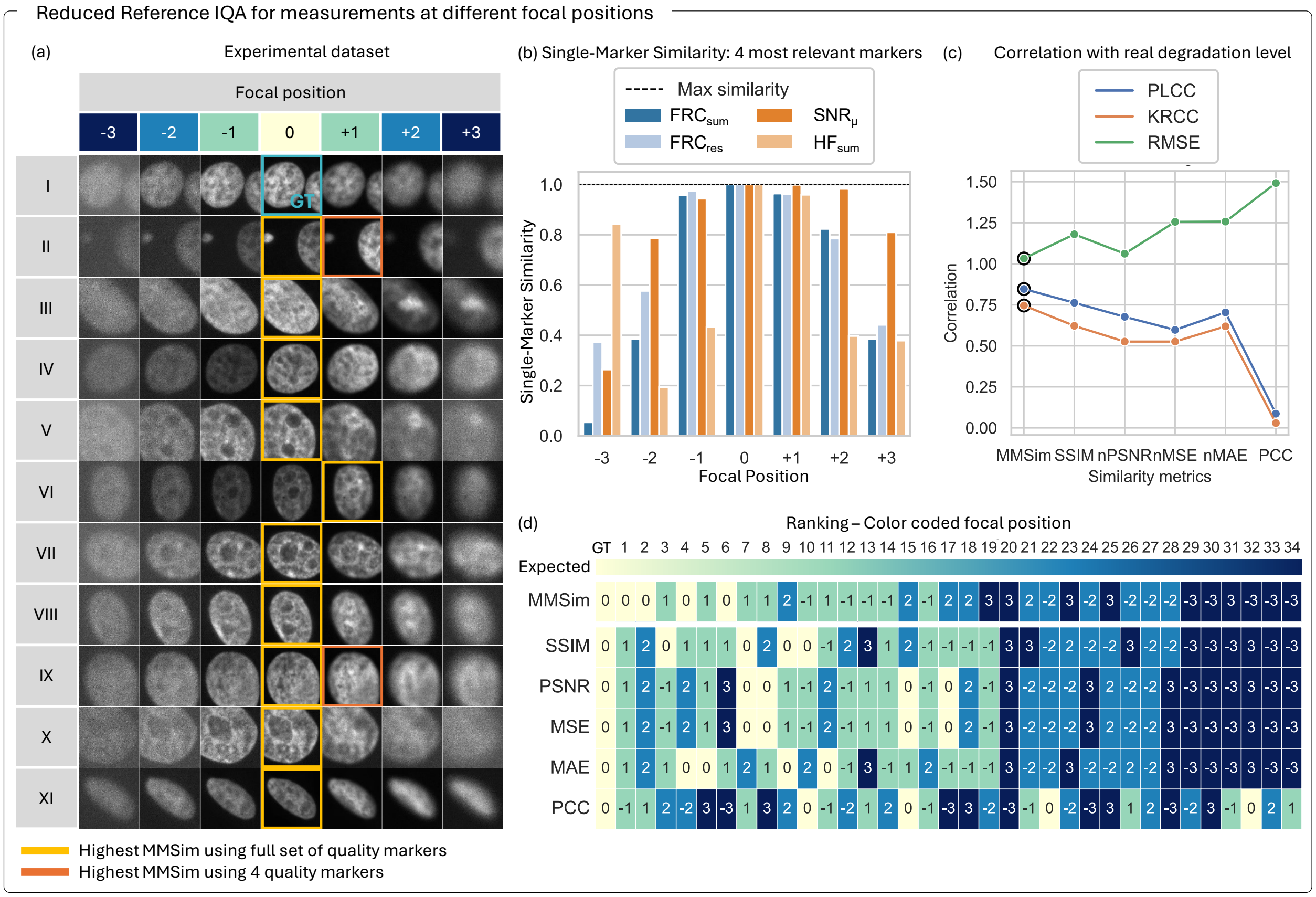

*Figure 6 – **Reduced-reference IQA for out-of-focus images.** (a) Eleven measurements of BPAE cells (rows) at different focal positions along the optical axis (columns). Position 0 is the optimal focal plane; positive and negative numbers indicate measurements above and below the optimal focal plane. Position 0 of FOV I is used as GT (blue frame). The yellow frames indicate the focal position of each FOV with the highest MMSim, computed using the full set of quality markers. The orange frames indicate the highest MMSim computed with the subset of the 4 most relevant markers (panel (b)), if the result differs from the one labelled by the yellow frame. (b) MMSim for the 4 most relevant markers automatically selected by principal component analysis, computed for FOV I. The GT, located in the centre, has the maximum similarity score for all quality markers, while the similarity decreases while moving from the optimal focal plane. (c) Correlation of MMSim and other FR metrics with the real image degradation. Black circles show that MMSim scores the highest PLCC and KRCC, and the lowest RMSE. (d) Global rankings for FOV I-V generated by different quality metrics (rows). The colour indicates the distance from the optimal focus, as indicated in panel (a). The coloured top bar shows the expected distribution, dependent only on the image quality: an ideal evaluation result would place the images at optimal focus (0, light yellow) at the top of the ranking, and sort the other images according to the distance from the optimal focus, placing sequentially ±1, ±2, and then ±3 positions.*

One drawback of our method is the slower computational time compared to other metrics, because MMSim requires the computation of multiple quality markers to generate the single-marker similarity scores. In the case of this dataset, the process must be repeated for seventy images (excluding the reference FOV), resulting in a long evaluation process. The set of NR metrics for the computation of the single-marker similarity scores have been selected to provide a comprehensive description of the images and to target a variety of experimental and processing artifacts. However, once the evaluation problem is clearly identified, not all metrics are equally relevant to IQA. Therefore, to reduce the complexity of the problem, we select the four most relevant metrics, i.e., the most variable with the defocusing process. The selection of the metrics can be applied after the manual inspection of the single-marker similarity computed for the reference FOV, but also in an automatic fashion, for example by principal component analysis (PCA) (see *Methods* section for further details). PCA is applied only to the reference FOV to ensure that the variability is not caused by the imaged objects but by the artifacts itself. Then, the metrics that contribute the most to the principal components are selected. In our case the most relevant metrics are, in order of relevance, FRC resolution, high frequency components, signal-to-average ratio, and FRC sum (Figure 6 (b)). By utilizing only this subset, the best focal positions are the same as those selected with the default set of metrics, except for FOV II and FOV IX, as indicated in Figure 6 (a) by the orange frames.

Figure 6 (c) shows three correlation measures between the IQA metrics and the real defocusing level: MMSim shows the best correlation due to a maximized Pearson linear correlation coefficient (PLCC) and Kendall rank correlation coefficient (KRCC) and a minimized root mean squared error (RMSE) compared to the other metrics. Exact correlation values are shown in supplementary Figure S5. Figure 6 (d) shows the overall quality ranking generated by all metrics for the first five FOVs coloured according to the focal position of the ranked measurements. Again, the interpretation of the ranking is supported by a coloured bar at the top of the panel, which displays the expected colour gradient for an ideal evaluation. MMSim places the optimal focal positions (0) and those close to the optimal focus (+1, -1) at top positions, following the expected trend, while other metrics, especially PCC, sort the image quality also according to the similarity between imaged objects.

# Discussion

In this manuscript, we have introduced MMSim, an evaluation method for estimating image similarity from the comparison of multiple image quality markers. The main difference with state-of-the-art FR approaches is the generation of a quality score that is not computed by pixel-wise or window-wise distances between the image and the GT but is based on the extraction and comparison of global quality markers. These markers include, but are not limited to, resolution, contrast, signal-to-noise ratio, structural complexity, and high frequency components.

MMSim achieves comparable performance with state-of-the-art full reference metrics when applied to FR IQA. In addition, it shows good agreement with visual perception, and high stability when applied to images acquired under the same experimental conditions. For example, for denoising applications MMSim favours images with low amount of residual noise rather than highly smoothed images, showing an opposite trend compared to state-of-the-art methods and selecting outputs with a higher level of detail. In fact, the indirect evaluation of the images through the extraction of global quality markers makes our method strongly dependent on the degradation content of the images and weakly dependent on the specific FOV, as long as the structure of the imaged objects remains similar. For this reason, MMSim offers the advantageous option to implement a RR evaluation, where a single FOV is used as reference for multiple measurements of the same kind of sample. This application provides a more flexible approach than the state-of-the-art.

Thanks to the definition of the quality markers, which are based on physical experimental artifacts and the resulting image degradations, MMSim can be further inspected by looking at the single-marker similarity scores. In this way, the result can be interpreted by revealing which markers affect the similarity with the GT and are most relevant for specific applications.

Compared to state-of-the-art FR metrics, the detailed image evaluation and the possibility to gain further insights come at the cost of higher computational time. Moreover, MMSim may fail when applied to image processing tasks that recover the desired quality features by introducing unexpected artifacts that can be only identified by locally inspecting the intensity values of individual pixels. To mitigate the first drawback, we took advantage of the interpretability of MMSim, and we tested an automatic approach for reducing the set of quality markers for specific evaluation problems. To address the second drawback, MMSim could be integrated with state-of-the-art evaluation metrics and used for a complementary IQA. In addition, MMSim could be used as a fine-tuning approach to select among good available options or to interpret the choice of other metrics.

In this study we have demonstrated the successful application of MMSim in a variety of scenarios: the hyperparameter optimization of denoising methods and the generation of large quality rankings in the presence of multiple processing results; the use of a RR implementation to generate quality rankings of noisy images and to optimize the focal position, and the successful simplification of the evaluation process by selecting a reduced set of metrics. Our method has proven to be stable, reliable and in good agreement with visual perception in all these cases. A reliable FR IQA approach is essential to validate experimental methods and processing algorithms, and to compare images to a known reference. In addition, the ability to interpret the evaluation in terms of individual image features gives further potential for inspecting degradation and processing outcomes to uncover unknown insights during a supervised FR IQA. Furthermore, the same potential can be transferred to a RR implementation, thus reducing the amount of prior information required and making the evaluation feasible for real experimental scenarios. For example, MMSim is feasible for an accurate, automatic and interpretable IQA for optical microscopy-based biomedical studies. Indeed, MMSim provides IQA, automatic generation of quality rankings and interpretation of a varied image collection. This is fundamental when assessing the measurement quality of experimental techniques, such as the noise content or blurring, and when applying processing algorithms for image analysis and diagnostic tasks.

# Methods

## Single-marker similarity and multi-marker similarity (MMSim)

Multi-marker similarity is based on a straightforward workflow comprising the following steps. Once the GT is defined for a given dataset:

1. A set of NR quality metrics are selected and computed for all images, including the GT. Each metric is considered as a quality marker for a specific image property, e.g., resolution, signal-to-noise ratio, contrast, and so on.
2. For each quality marker, a single marker similarity is constructed to evaluate the similarity between the images and the GT according to a single image property.
3. The single-marker similarities are averaged to generate a multi-marker similarity (MMSim) that provides an estimate of the global similarity of the images and the GT considering all the selected quality markers.

In the following, we provide the definitions for all the single steps explained here.

***Single quality markers***

The NR metrics are selected to provide a comprehensive IQA of the dataset. They should address complementary aspects of the images because one single metric is usually not sufficient to detect multiple image degradations caused by experimental and processing algorithms. Inspired by our previous study[19], we selected eight NR metrics.

The definitions of the NR metrics are given below for an image *s*. In the following expressions $N_{pix}$ is the number of pixels of the image, $s_{max}$ the maximum value of the image, $\mu_s$ and $\sigma_s$ the average and the standard deviation of the image or, if available, of a selected background region.

- *Fourier Ring Correlation (FRC)*. FRC is a correlation-based measure of the spectral signal-to-noise of an image, and it is commonly used to quantify the image resolution. FRC is computed by correlating rings of increasing radius extracted from two independent versions of the image in the Fourier domain. In this work, we the single-image implementation of FRC[21] and we utilized as markers both the estimated resolution ($FRC_{res}$) and the sum of the correlation curve ($FRC_{sum}$).
- *Signal to noise ratio* $SNR_\sigma$, defined as the logarithmic ratio between the dynamic range of the signal and the standard deviation of a background region. If the background region is not selected a priori, the full image is used to compute the standard deviation.
$$SNR_\sigma(s) = 20\log_{10}\frac{s_{max}-\mu_s}{\sigma_s}$$
- *Contrast to average ratio* $SNR_\mu$, defined as the logarithmic ratio between the dynamic range of the signal and the mean value of a background region. If the background region is not selected a priori, the full image is used to compute the mean value.
$$SNR_\mu(s) = 20\log_{10}\frac{s_{max}-\mu_s}{\mu_s}$$
- *Sum of high frequency components* $HF_{sum}$, defined as the sum of all pixels of the sum-normalized image power spectrum lying outside a radial threshold. The radial threshold selects the high-frequency part of the power spectrum, and it is set to 3/8 the full size of the image.
$$HF_{sum}(s) = \sum_{f_x^2+f_y^2>f_{th}^2}\frac{|\mathcal{F}(s)|^2}{\sum_{N_{pix}}|\mathcal{F}(s)|^2}$$
- *Structural complexity SC*, defined as the average gradient of the image. It accounts for the presence of sharp structures or noise, and its value is lower for smoothed of flat intensity profile images.
$$SC(s) = \frac{1}{N_{pix}}\sum_{N_{pix}}\sqrt{\frac{\nabla_x^2(s)+\nabla_y^2(s)}{2}}$$
- *Edge energy ratio* $E_{edge}$. It is the ratio between the sum of the pixels lying outside a radial threshold and the sum of all the pixels of the image. In presence of vignetting, $E_{edge}$ is lower. The threshold is set to 3/8 the full size of the image.
$$E_{edge}(s) = \frac{\sum_{x^2+y^2>r_{th}^2} s(x,y)}{\sum_{x,y} s(x,y)}$$
- *Root mean square contrast* $C_{RMS}$, that provides an additional contrast definition, and it can be affected also by the presence of noise. The image is rescaled between 0 and 1 before computation of $C_{RMS}$.
$$C_{RMS} = \sqrt{\frac{1}{N_{pix}}\sum_{N_{pix}}(s-\mu_s)^2}$$

The quality markers assume always positive values, and the possible negative values reached for the logarithmic SNRs can be regularized to a small constant ε. Prior computation of the metrics, the images are rescaled to in the interval [0,1]. In this manuscript $\mu_s$ and $\sigma_s$ are computed for the full image. Once the markers are computed for the whole dataset, strong outliers that may be generated due to presence of strong artifacts are removed by locking the minimum and maximum value of the markers to the lower percentile of 0.005 and upper percentile of 0.995. Such normalizations are recommended for a large dataset and must be computed consistently for the entire set of images but may be redundant and not necessary for simpler cases.

***Definition of single-marker similarity***

Given an image $s$ and the GT $s_{GT}$, the single-marker similarity (SMSim) is built to quantify their similarity with a quantitative score. SMSim should be symmetric, i.e., $SMSim\ (s, s_{GT}) = SMSim\ (s_{GT}, s)$, and bounded in a certain interval independently on the order of magnitude of the quality marker. Various analytical functions satisfy these conditions, and they are characterized by different curves. We have selected for the SMSim an expression that resembles the one for the single factors in the structural similarity index (SSIM). the SMSim for a marker $m_i$ is defined as:

$$SMSim_i(s, s_{GT}) = \frac{2m_{i,s}m_{i,s_{GT}} + k}{m_{i,s}^2 + m_{i,s_{GT}}^2 + k}$$

SMSim is 1 if the marker assumes the same value for the two images, and decreases, with a lower limit of 0, if the two markers assume different values. The small constant $k$ is introduced to regularize the result in case of very small values of $m_{i,s}$ and $m_{i,sGT}$, and it is equal to 0.01 the data range of the marker $m_i$. Before computation of SMSim, the metrics are rescaled to the range [0,1]. Therefore, the data range of all markers is 1 and $k = 0.01$.

***Definition of multi-marker similarity***

MMSim is the global similarity score which accounts for all the single-marker similarity scores computed for the images and it should preserve the bound of the single score. Among the averaging options available, we choose a simple arithmetic average, which provides same importance to all markers. For a set of single-marker similarity scores $\text{SMSim}_i$ computed on $N_m$ markers $m_i$ for an image $s$ and the GT $s_{GT}$:

$$MMSim(s, s_{GT}) = \frac{1}{N_m}\sum_{m_i} SMSim_i(s, s_{GT})$$

MMSim is bound to the interval [0,1] and its value is higher for better similarity of the images.

***FR state-of-the-art metrics***

MMSim is benchmarked to state-of-the-art FR metrics commonly used by the imaging community. In particular, we compare our method with:

- Structural similarity index:
$$SSIM(s, s_{GT}) = \frac{(2\mu_s\mu_{s_{GT}} + C_1)(2\sigma_{s,s_{GT}} + C_2)}{(\mu_s^2 + \mu_{s_{GT}}^2 + C_1)(\sigma_s^2 + \sigma_{s_{GT}}^2 + C_2)}$$
With $C_1=(k_1L)^2$ and $C_2=(k_2L)^2$, $k_2 = 0.01$, $k_2 = 0.03$, and L equal to the range value of the image pixels.
- Mean squared error:
$$MSE(s, s_{GT}) = \frac{1}{N_{pix}}\sum_{i=1}^{N_{pix}}(s_i - s_{GT,i})^2$$
- Peak signal-to-noise ratio, defined as:
$$PSNR(s, s_{GT}) = 20\log_{10}\frac{s_{GT,max}}{\sqrt{MSE(s, s_{GT})}}$$

- Mean absolute error:

$$MAE(s, s_{GT}) = \frac{1}{N_{pix}} \sum_{i=1}^{N_{pix}} |s_i - s_{GT,i}|$$

- Pearson's correlation coefficient:

$$PCC(s, s_{GT}) = \frac{\sum_{i=1}^{N_{pix}} (s - \bar{s})(s_{GT} - \overline{s_{GT}})}{\sqrt{\sum_{i=1}^{N_{pix}} (s - \bar{s})^2} \sqrt{\sum_{i=1}^{N_{pix}} (s_{GT} - \overline{s_{GT}})^2}}$$

***Workflow for IQA and quality ranking***

IQA is executed by computing MMSim and additional FR metrics on the entire dataset, upon selection of the GT. The quality rankings are generated by sorting the images from highest to lowest score for each metric. The scores are also computed for the GT itself, which is placed at the top of the ranking with a value of 1. Visual comparison of the metrics can be facilitated by plotting the results. This approach requires the scores to be constrained to the same interval and follow the same trend (increasing or decreasing) for an improving similarity. We decided to standardize all metrics to have an increasing score up to 1 for an improving similarity. MMSim, SSIM and PCC already satisfy these requirements. PSNR already shows the desired behaviour but assumes an infinite value for perfect similarity. Therefore, the infinite value is set to 110% of the highest score of the dataset. Then, percentile normalization is applied with minimum quantile of 0.05 and maximum quantile of 0.95 to avoid strong outliers for visualization. MSE and MAE show a decreasing behaviour with increasing similarity and have value 0 for perfect similarity. Therefore, their value is first clipped between 0.05 and 0.95 quantiles, then their trend is inverted. For the metric x with value 0 for perfect similarity, the inverted trend $x_{inv}$ is obtained by calculating $x_{inv}=x_{min}/x$, here $x_{min}$ is the minimum value in the dataset. Normalized values of PSNR, MSE and MAE are indicated in this manuscript, respectively, as nPSNR, nMSE, and nMAE.

***Reduction of the set of quality markers by principal component analysis (PCA)***

To reduce the computational and time effort for specific applications, we integrated few additional steps in the evaluation workflow. This approach is particularly feasible for RR IQA and to identify which metrics are strongly affected by the image degradations. Initially, the GT is selected among a subset of images of the same FOV, i.e., images with different quality but the same content. Then, the full set of quality markers is computed for these images, and the markers are used to fit a PCA model before any estimation of similarity. Then, the PCA components are inspected by extracting their coefficients and explained variance. The *N* most relevant metrics are selected by rescaling their corresponding coefficients according to the explained variance ratio and summing the resulting values over the set of components. In this manuscript, we selected the 4 markers with highest relevance. Finally, the selected metrics are used for the computation of single-marker similarities and MMSim for the entire dataset, including more FOVs and implementing a RR evaluation. This process can be adapted by selecting the number of relevant metrics to be extracted or a threshold for their relevance.

# Datasets

***Experimental images with different noise levels: Fluorescence Microscopy Denoising (FMD) dataset***

We utilized confocal microscopy measurements of fixed bovine pulmonary artery endothelial (BPAE) cells and fixed zebrafish embryos from the FMD dataset.[20] The measurements are provided with six noise levels, obtained by averaging an increasing number of raw measurements. The noisy images are generated by averaging 1, 2, 4, 8, and 16 raw images, while the high-quality images are obtained from the average of 50 raw images. The measurements of BPAE cells are composed by three channels: nuclei labelled with DAPI (blue), F-actin labelled with Alexa Fluor 488 phalloidin (green), and mitochondria labelled with MitoTracker Red CMXRos (red). The measurements of zebrafish at 2 days post fertilization are composed by a single channel. These experimental images are utilized to evaluate the similarity with increasing levels of noise (Figure 4 and Figure 5), and for denoising applications (Figure 2 and Figure 3).

### *Hyperparameter optimization for denoising by Gaussian filtering*

The study on the optimization of Gaussian filtering is implemented by iteratively smoothing the image with a bidimensional Gaussian function of increasing standard deviation σ, as reported in Figure 2 (c). The iterative process is repeated for each FR metric. The standard deviation of the first iteration is set to $\sigma_1 = 0.05$ px, then it is increased at each iteration by $\delta\sigma = 0.05$ px. At each iteration, the value of the FR metric is checked, and the process is continued until the value is improved. If an optimization is not obtained within 100 iterations, the process is stopped. The improvement is checked by monitoring the relative increase (for MMSim, SSIM, PSNR) or decrease (for MSE and MAE) between two consecutive iterations, with a tolerance of 1e-5 of the value assumed at the previous iteration, to avoid early stopping due to small oscillations of the metrics.

### *IQA and quality ranking of denoised images*

The study on different denoising methods utilizes raw images of the FMD datasets denoised with the following methods: average filter, median filter (SciPy implementation in python), Gaussian filter, Total Variation regularization (Scikit-image implementation in python), Wavelet filtering (Scikit-image implementation in python). The GT is selected among the high-quality images obtained by averaging 50 raw images. The hyperparameters utilized for the generation of the denoised images of Figure 3 (b) are included in Table 1.

| Denoising method | Hyperparameter | P1 | P2 | P3 | P4 | P5 |
|---|---|---|---|---|---|---|
| Average filter | Lateral filter size / px | 3 | 5 | 7 | 9 | 11 |
| Gaussian filter | Standard deviation of 2D Gaussian function / px | 1 | 2 | 3 | 4 | 5 |
| Median filter | Lateral filter size / px | 3 | 5 | 7 | 9 | 11 |
| Total Variation regularization | Denoising weight | 0.01 | 0.1 | 0.5 | 1 | 2 |
| Wavelet filtering (db1) | Noise standard deviation | 10 | 20 | 30 | 40 | 50 |

*Table 1 – Denoising methods and relative hyperparameters utilized to evaluate experimental denoised images.*

### *Experimental images at different focal planes*

Experimental images at different focal positions are collected from the open-source dataset provided by C. Zhang et al.[22] We have selected eleven confocal measurements of BPAE cells, acquired at seven focal positions: at the optimal focal plane and at steps of 0.6, 1.2 and 1.8 µm along the z-axis, below and above the ideal focus. These images are utilized to demonstrate the applicability of MMSim for defocusing problems and to prove its feasibility for RR implementations.

## Validation of metrics through correlation measures

The performance of MMSim against state-of-the-art metrics is evaluated by correlating each IQA method with the real image degradation through Pearson linear correlation coefficient (PLCC), Kendall rank correlation coefficient (KRCC), and root mean squared error (RMSE). To compute this evaluation, we use measurements with known experimental artifacts, for which it is possible to sort the images objectively according to their quality. Depending on the strength of the artifact, we assign an integer number that represents a quality score.

A score from 0 to 3 is assigned to the out-of-focus images. The score of 0 is assigned to the positions furthest from the optimal focus, while the score of 3 is assigned to the in-focus images. In the case of images exhibiting increasing noise (Figure 4), we test two approaches: in the first one, a score from 1 to 6 is assigned to the images of generated using an increasing number of averaged images, while in the second configuration, the score is the number of images averaged to obtain the measurement (1, 2, 4, 8, 16, and 50). Finally, these quality scores are correlated with the similarity predicted by each metric. The results are reported in supplementary Figure S3, S4 and S5.

# Acknowledgments

**Author contributions**: E.C. performed the study, E.C. and T.B. conceptualized the idea and prepared the manuscript, T.B. supervised the overall research study.

**Funding**: This work is supported by the BMBF, funding program Photonics Research Germany [13N15706 (LPI-BT2-FSU), 13N15719 (LPI-BT5)] and is integrated into the Leibniz Center for Photonics in Infection Research (LPI). The LPI initiated by Leibniz-IPHT, Leibniz-HKI, Friedrich Schiller University Jena and Jena University Hospital is part of the BMBF national roadmap for research infrastructures. The work presented has received funding from the European Union's Horizon 2020 research and innovation programme [grant agreement No. 860185 (PHAST)]. Co-funded by the European Union [ERC, STAIN-IT, 101088997]. Views and opinions expressed are however those of the author(s) only and do not necessarily reflect those of the European Union or the European Research Council. Neither the European Union nor the granting authority can be held responsible for them.

**Competing interests**: The authors declare that there is no conflict of interest regarding the publication of this article.

# Data availability

The source code for Multi-Marker Similarity is publicly available in the GitLab repository https://git.photonicdata.science/elena.corbetta/multi-marker-similarity with a small demo dataset. The images used in this study are taken or generated from open-source datasets. Source images and source data for the reproduction of the results of this study are available from the authors upon reasonable request.

# Supplementary information

## Optimization of denoising results of multi-channel images

MMSim and state-of-the-art FR metrics are utilized to predict the best combination of denoising results for RGB images of BPAE cells. In Figure S1, noisy and GT images of 5 different FOVs are compared to the optimized images selected by the quality metrics. Each channel is denoised and evaluated independently, then the best denoised channels for each FOV is predicted by the metrics, generating the multi-channel images by combining the best denoised image of each channel. In general, all the metrics show a good result, even if PCC is biased towards higher levels of smoothing. MMSim recovers well the image features and selects images with similar quality, independently on the FOV. PSNR, MSE and MAE show also a stable performance. SSIM selects images with good quality, with a bias towards stronger smoothing for the green and blues channel.

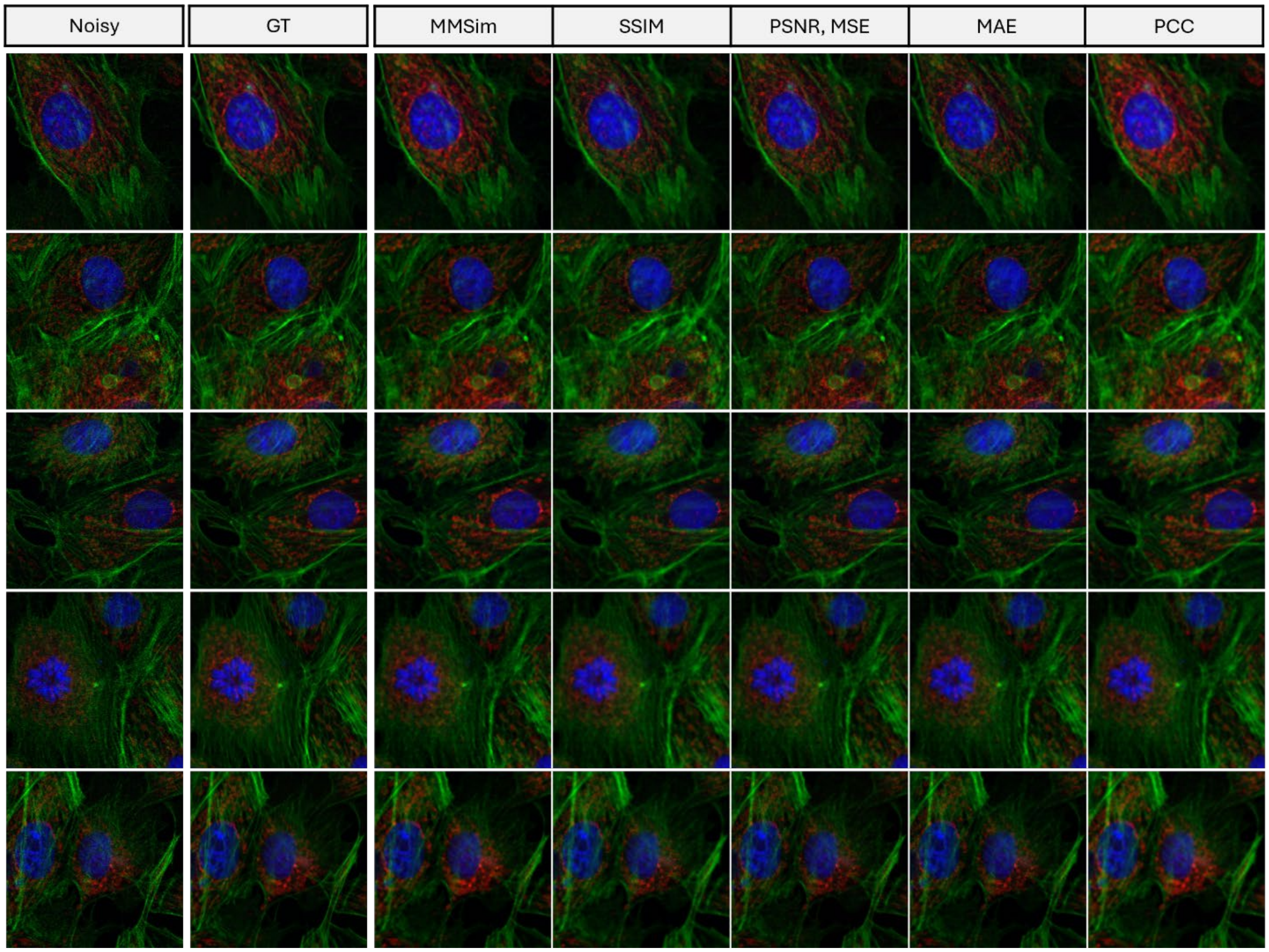

*Figure S1 – **Optimization of 3-channel measurements of BPAE cells.** Noisy images are denoised by 5 denoising methods (average filter, Gaussian filter, median filter, total variation regularization, and wavelet filtering), applied with 5 values of the hyperparameters, as in Figure 3 of the main manuscript. The plot shows the GT images, obtained by averaging 50 noisy measurements, and the optimized combination of denoised channels as predicted by MMSim and other FR metrics. All metrics show a good performance, except for PCC, that selects images with excessive smoothing. MMSim shows a good recovery of all channels, selecting images with a similar quality across different measurements. SSIM selects more smoothed images for the green and blue channels and weaker smoothing for red channel.*

## Single marker similarity (SMSim) for experimental images denoised by different methods

Confocal measurements of fixed zebrafish embryos are denoised by different methods, for which one hyperparameters is tuned over 5 different values. Then, SMSim is computed for the set of 8 quality markers and is plotted in Figure S2. The plots enable further inspection of the similarity scores. In case of this denoising application, the presence of noise or excessive smoothing affects the similarity in terms of high frequency components ($HF_{sum}$) and structural complexity ($SC$). However, few methods strongly modify additional quality features of the images: the median filter reduces the similarity given by the signal-to-average ratio ($SNR_{\mu}$), while the wavelet filtering affects the Fourier ring correlation estimate ($FRC_{sum}$). The combination of these trends determines the final multi-marker similarity (MMSim) score reported in Figure 3.

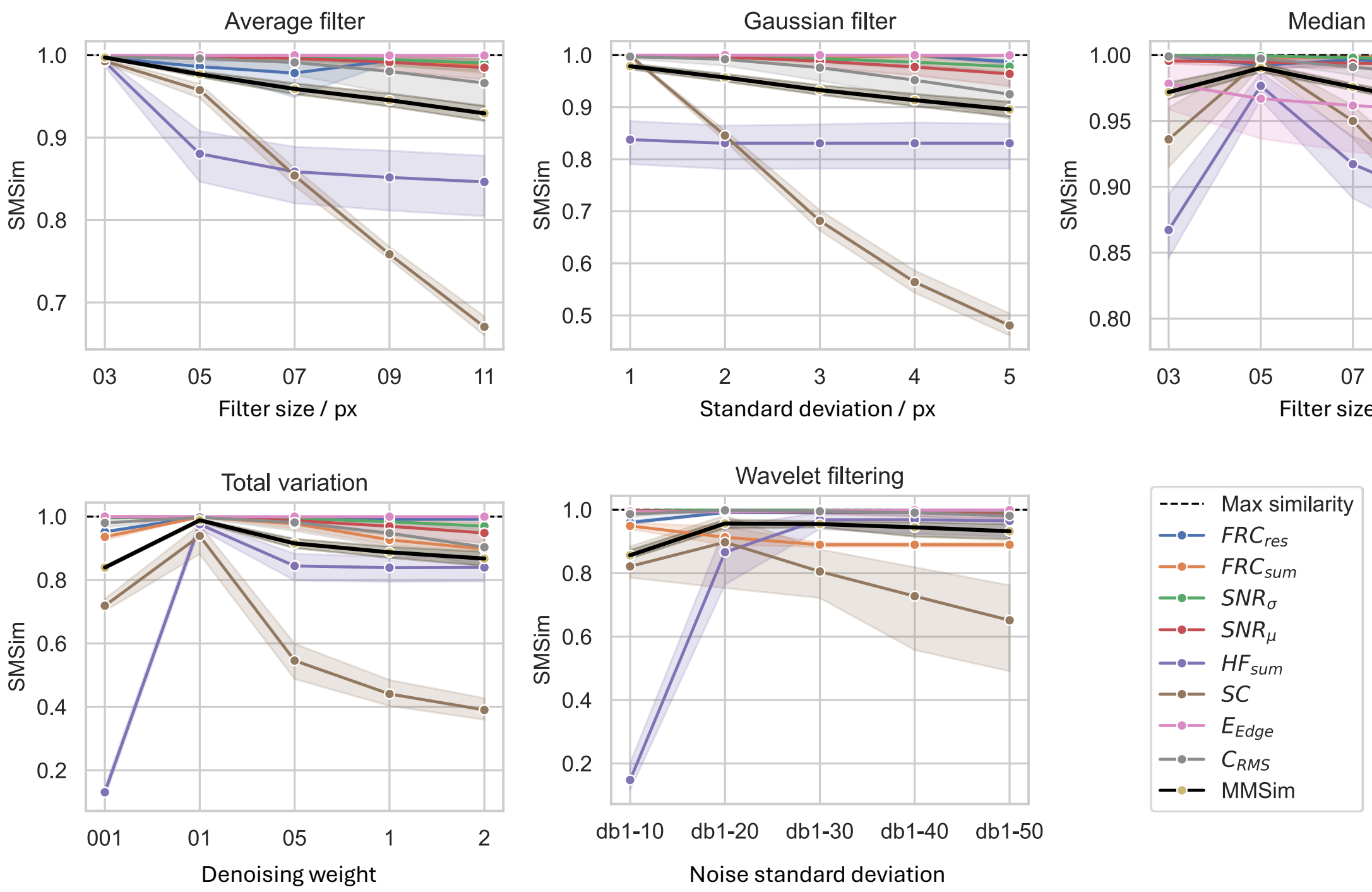

*Figure S2 – Single marker similarity (SMSim) computed for denoised images. The title of each plot refers to the denoising methods, while the x axis indicates the value of the hyperparameter.*

## Correlation between real degradation level and FR metrics

The performance of MMSim is validated against the state-of-the-art metrics. We computed the Pearson linear correlation coefficient (PLCC), the Kendall rank correlation coefficient (KRCC), and the root mean squared error (RMSE) for all the full-reference (FR) metrics. The metrics are correlated with objective scores assigned to the images according to the real experimental degradation. The scores have higher values for increasing image quality. A good result is obtained by maximizing PLCC and KRCC results and by minimizing the RMSE.

Figure S3 shows the correlation results the confocal measurements of fixed zebrafish embryos with increasing noise levels. The correlation is computed in a FR configuration, selecting one GT for each field of view, and it is repeated for two different objective scores: in the first case, the scores are increasing integer numbers from 0 to 6 (Figure S3 (a)), while in the second case they correspond to the number of images averaged to obtain each noise level (Figure S3 (b)). In this case, our method shows good correlation results, reaching values of PLCC, KRCC and RMSE very close to the best ones. MMSim shows poor correlation only for PLCC in panel (b); nevertheless, it reaches good values of KRCC and RMSE in the same panel, demonstrating that it is comparable or better than established state-of-the-art approaches.

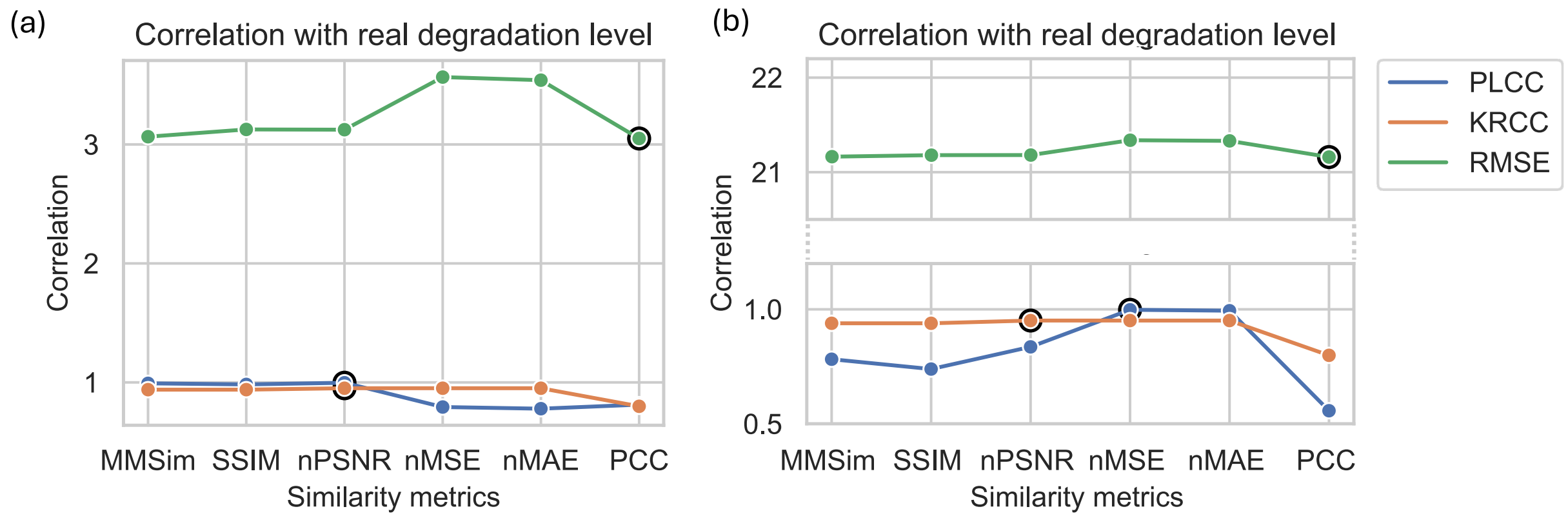

| Similarity metrics | PLCC | KRCC | RMSE |
|---|---|---|---|
| **MMSim** | 0,992 | 0,939 | 3,065 |
| **SSIM** | 0,983 | 0,939 | 3,126 |
| **nPSNR** | 0,997 | 0,951 | 3,124 |
| **nMSE** | 0,793 | 0,951 | 3,566 |
| **nMAE** | 0,779 | 0,951 | 3,540 |
| **PCC** | 0,813 | 0,799 | 3,050 |

| Similarity metrics | PLCC | KRCC | RMSE |
|---|---|---|---|
| **MMSim** | 0,783 | 0,939 | 21,164 |
| **SSIM** | 0,739 | 0,939 | 21,180 |
| **nPSNR** | 0,836 | 0,951 | 21,181 |
| **nMSE** | 0,998 | 0,951 | 21,338 |
| **nMAE** | 0,994 | 0,951 | 21,331 |
| **PCC** | 0,557 | 0,799 | 21,160 |

*Figure S3 – Correlation measures computed for five confocal measurements of fixed zebrafish embryos with different noise levels. Similarity metrics computed in a FR configuration. (a) The score assigned to the images is an integer number from 0 (raw image) to 6 (high-quality image), increasing with the image quality. (b) The score corresponds to the number of images averaged to obtain each noise level. They are, respectively, from the raw image to the high-quality image: 1, 2, 4, 8, 16, 50. The black round frames in the plots indicate the best correlation result among the metrics.*

Figure S4 shows the correlation results for the same images, but in the reduced-reference (RR) configuration, as reported in the main manuscript (Figure 4). MMSim confirms its feasibility for RR evaluations, reaching the best correlation results for PLCC and KRCC, and minimizing the RMSE. Figure S5 shows the correlation results obtained for images at different focal positions in a reduced reference configuration (Figure 6). In both cases, MMSim shows the best result, demonstrating the reliability of the method and its ability to track the changes induced by the image degradations by following the changes in the quality markers. In both cases, the evaluation is executed in a reduced-reference configuration.

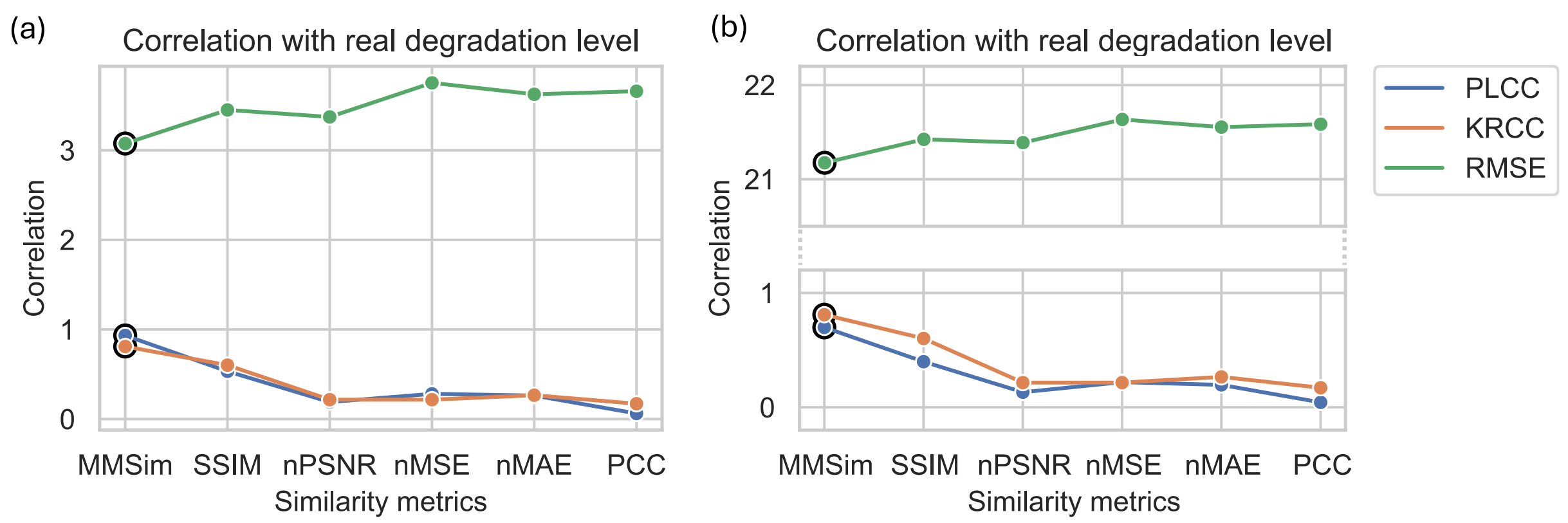

| Similarity metrics | PLCC | KRCC | RMSE |
|---|---|---|---|
| **MMSim** | 0,926 | 0,785 | 3,078 |
| **SSIM** | 0,533 | 0,602 | 3,451 |
| **nPSNR** | 0,190 | 0,216 | 3,373 |
| **nMSE** | 0,281 | 0,216 | 3,753 |
| **nMAE** | 0,262 | 0,266 | 3,626 |
| **PCC** | 0,062 | 0,171 | 3,660 |

| Similarity metrics | PLCC | KRCC | RMSE |
|---|---|---|---|
| **MMSim** | 0,695 | 0,785 | 21,176 |
| **SSIM** | 0,399 | 0,602 | 21,424 |
| **nPSNR** | 0,132 | 0,216 | 21,389 |
| **nMSE** | 0,220 | 0,216 | 21,635 |
| **nMAE** | 0,196 | 0,266 | 21,554 |
| **PCC** | 0,043 | 0,171 | 21,585 |

*Figure S4 – Correlation measures computed for five confocal measurements of fixed zebrafish embryos with different noise levels. Similarity metrics computed in a RR configuration. (a) The score assigned to the images is an integer number from 0 (raw image) to 6 (high-quality image), increasing with the image quality. (b) The score corresponds to the number of images averaged to obtain each noise level. They are, respectively, from the raw image to the high-quality image: 1, 2, 4, 8, 16, 50. The black round frames in the plots indicate the best correlation result among the metrics, obtained always by MMSim.*

| Similarity metrics | PLCC | KRCC | RMSE |
|---|---|---|---|
| **MMSim** | 0,847 | 0,746 | 1,033 |
| **SSIM** | 0,762 | 0,622 | 1,180 |
| **nPSNR** | 0,677 | 0,526 | 1,061 |
| **nMSE** | 0,598 | 0,526 | 1,256 |
| **nMAE** | 0,704 | 0,619 | 1,257 |
| **PCC** | 0,086 | 0,029 | 1,492 |

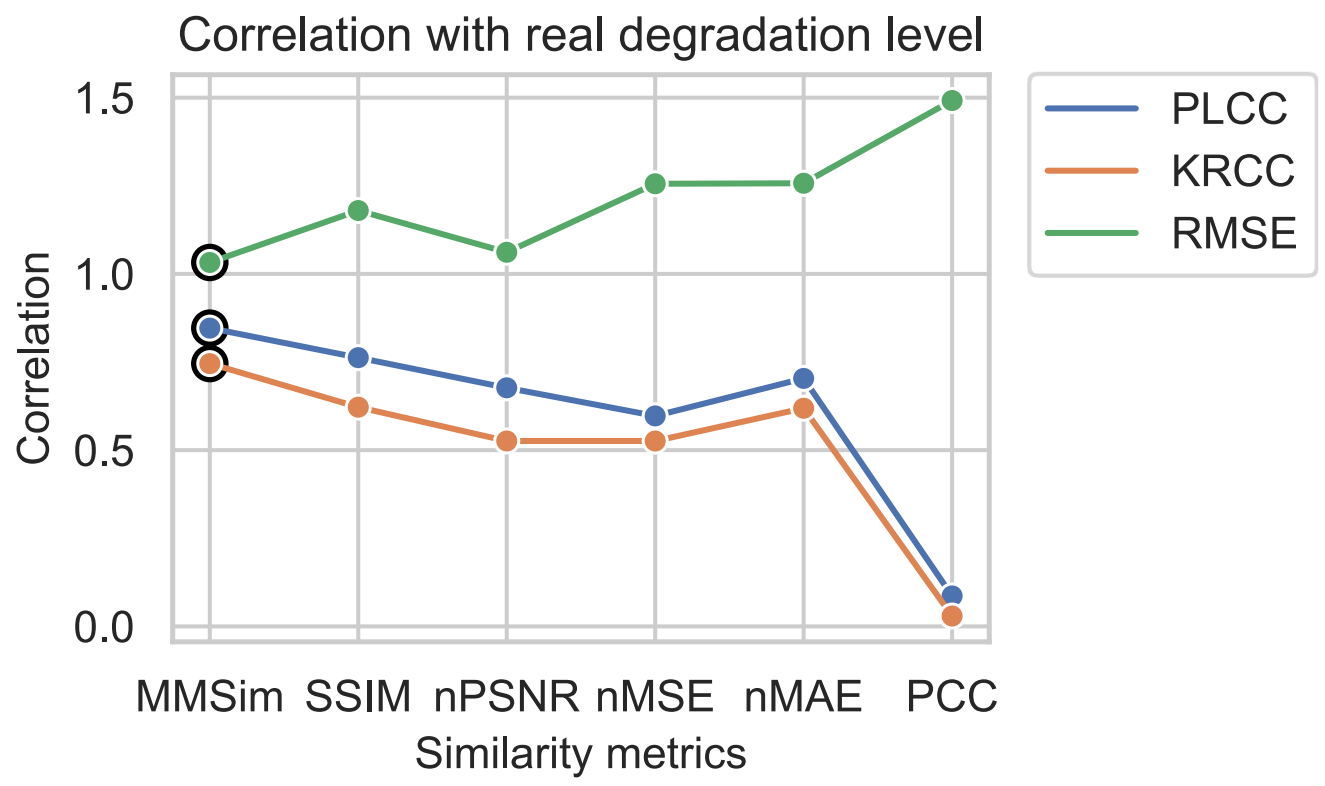

*Figure S5 – Correlation measures computed for eleven confocal measurements of BPAE cells at seven different focal positions along the optical axis. Similarity metrics computed in a FR configuration. The score assigned to the images is an integer number from 0 (strongest defocusing) to 3 (optimal focal plane). The black round frames in the plots indicate the best correlation result among the metrics, obtained always by MMSim.*